\documentclass[fleqn,usenatbib,useAMS]{mnras}

\usepackage{newtxtext,newtxmath}
\usepackage[T1]{fontenc}

\DeclareRobustCommand{\VAN}[3]{#2}
\let\VANthebibliography\thebibliography
\def\thebibliography{\DeclareRobustCommand{\VAN}[3]{##3}\VANthebibliography}

\usepackage{graphicx}	% Including figure files
\usepackage{amsmath}	% Advanced maths commands
\usepackage{amssymb}	% Extra maths symbols

\title[Gravitational-wave localization with lensing]{Localizing merging black holes with sub-arcsecond precision using gravitational-wave lensing}

\author[O. A. Hannuksela et al.]{
Otto A. Hannuksela,$^{1,2}$\thanks{E-mail: o.hannuksela@nikhef.nl}
Thomas E. Collett,$^{3}$\thanks{E-mail: thomas.collett@port.ac.uk}
Mesut \c{C}al{\i}\c{s}kan$^{4}$
and Tjonnie G. F. Li$^{5}$
\\
$^{1}$Nikhef -- National Institute for Subatomic Physics, Science Park, 1098 XG Amsterdam, The Netherlands\\
$^{2}$Department of Physics, Utrecht University, Princetonplein 1, 3584 CC Utrecht, The Netherlands\\
$^{3}$Institute of Cosmology and Gravitation, University of Portsmouth, Burnaby Rd, Portsmouth, PO1 3FX, UK\\
$^{4}$Department of Astronomy and Astrophysics, and Kavli Institute for Cosmological Physics, University of Chicago, Chicago, IL 60637, USA\\
$^{5}$Department of Physics, The Chinese University of Hong Kong, Shatin, NT, Hong Kong
}

\pubyear{2020}

\begin{document}
\label{firstpage}
\pagerange{\pageref{firstpage}--\pageref{lastpage}}
\maketitle

% Abstract of the paper
\begin{abstract}
The current gravitational-wave localization methods rely mainly on sources with electromagnetic counterparts. Unfortunately, a binary black hole does not emit light. Due to this, it is generally not possible to localize these objects precisely. However, strongly lensed gravitational waves, which are forecasted in this decade, could allow us to localize the binary by locating its lensed host galaxy. Identifying the correct host galaxy is challenging because there are hundreds to thousands of other lensed galaxies within the sky area spanned by the gravitational-wave observation. However, we can constrain the lensing galaxy's physical properties through both gravitational-wave and electromagnetic observations. We show that these simultaneous constraints allow one to localize quadruply lensed waves to one or at most a few galaxies with the LIGO/Virgo/Kagra network in typical scenarios. Once we identify the host, we can localize the binary to two sub-arcsec regions within the host galaxy. Moreover, we demonstrate how to use the system to measure the Hubble constant as a proof-of-principle application. 
\end{abstract}

% Select between one and six entries from the list of approved keywords.
% Don't make up new ones.

%%%%%%%%%%%%%%%%%%%%%%%%%%%%%%%%%%%%%%%%%%%%%%%%%%

%%%%%%%%%%%%%%%%% BODY OF PAPER %%%%%%%%%%%%%%%%%%

\section{Introduction}

With current gravitational-wave (GW) detectors, the sky localization areas of GW events have typical uncertainties of 100s of square degrees \citep{LIGOScientific:2018mvr}. There are $>$millions of galaxies in such a large sky area, and tens of thousands of galaxies within the 90\% error volume~\citep{Chen:2017rfc,Fishbach:2018gjp,Gray:2019ksv,Abbott:2019yzh,Soares-Santos:2019irc}, making identification of the GW event host galaxy impossible unless there is an electromagnetic (EM) counterpart. This  was the case for the binary neutron star GW170817 \citep{TheLIGOScientific:2017qsa,Abbott:2017xzu}.

A binary black hole merger would allow us to probe physics inaccessible with merging neutron stars such as higher-order GW modes, higher source redshifts, and the strong field of gravity \cite{Berti:2015itd,Bustillo:2016gid,Pang:2018hjb,Chatziioannou:2019dsz,LIGOScientific:2019fpa,Miller:2019vfc,LIGOScientific:2020stg}. Unfortunately, localizing merging black holes is difficult as they emit no light\footnote{See, however, the possibility of identifying ``golden binaries'' that could allow for a unique localization~\cite{Chen:2016tys}.}. In this work, we ask the question: if a GW is multiply imaged due to strong gravitational lensing, would that allow us to uniquely localize the event through locating its multiply imaged host galaxy? 

Similarly to light, when GWs travel near a massive object such as a galaxy or a galaxy cluster, they experience gravitational lensing. These lensed GWs could be observed in this decade: The current single detector forecasts predict around one strongly lensed event per year at LIGO design sensitivity \cite{2018MNRAS.476.2220L,PhysRevD.97.023012,2018MNRAS.480.3842O}. The methods to detect lensed waves have been developed in recent years, and the first searches for gravitational-wave lensing signatures in the LIGO and Virgo data were carried out recently~\cite{2018arXiv180707062H,Hannuksela_2019,Li:2019osa,McIsaac:2019use,Pang:2020qow,2020arXiv200712709D}. 

If a GW event is gravitationally strongly lensed, then its host galaxy must also be lensed. Therefore, when we look for the host galaxy of a GW, we can narrow down our search to strongly lensed galaxies only. Given that there are far fewer strongly lensed galaxies than non-lensed galaxies \citep{collett2015}, this means that the number of possible hosts is orders of magnitude smaller compared to non-lensed GWs.

When gravitational lensing produces multiple images, typically either two or four bright images form (although in rare scenarios, more images are possible \cite{collettbacon2016,Dahle:2012bd,Collett:2017ksf}). Because the multiple images of the wave travel on different paths through the Universe, images of transient events do not reach Earth simultaneously. 
Therefore, a GW detector observes multiple images as "repeated" events with an overall difference in amplitude and possibly phase, separated by typically time delays of minutes to months \citep{2018arXiv180707062H,Hannuksela_2019,Smith:2017mqu,smith2019discovery,Robertson:2020mfh}. 
In the limit of geometrical optics, the gravitational wave is otherwise identical to a non-lensed signal.\footnote{Let us note that Ref.~\cite{Dai:2017huk} suggested that if the GW contains higher-order modes and passes through a lensing saddle point, the signal morphology could exhibit a minor change. Moreover, if beating patterns induced by microlensing are present, there could be minor changes to the waveform in the case of extreme macromodel magnification~\cite{diego2019observational,Pagano:2020rwj}. However, the effects are rare, and are not expected to significantly affect parameter estimation. }

\begin{figure}
    \centering
    \includegraphics[width=\columnwidth]{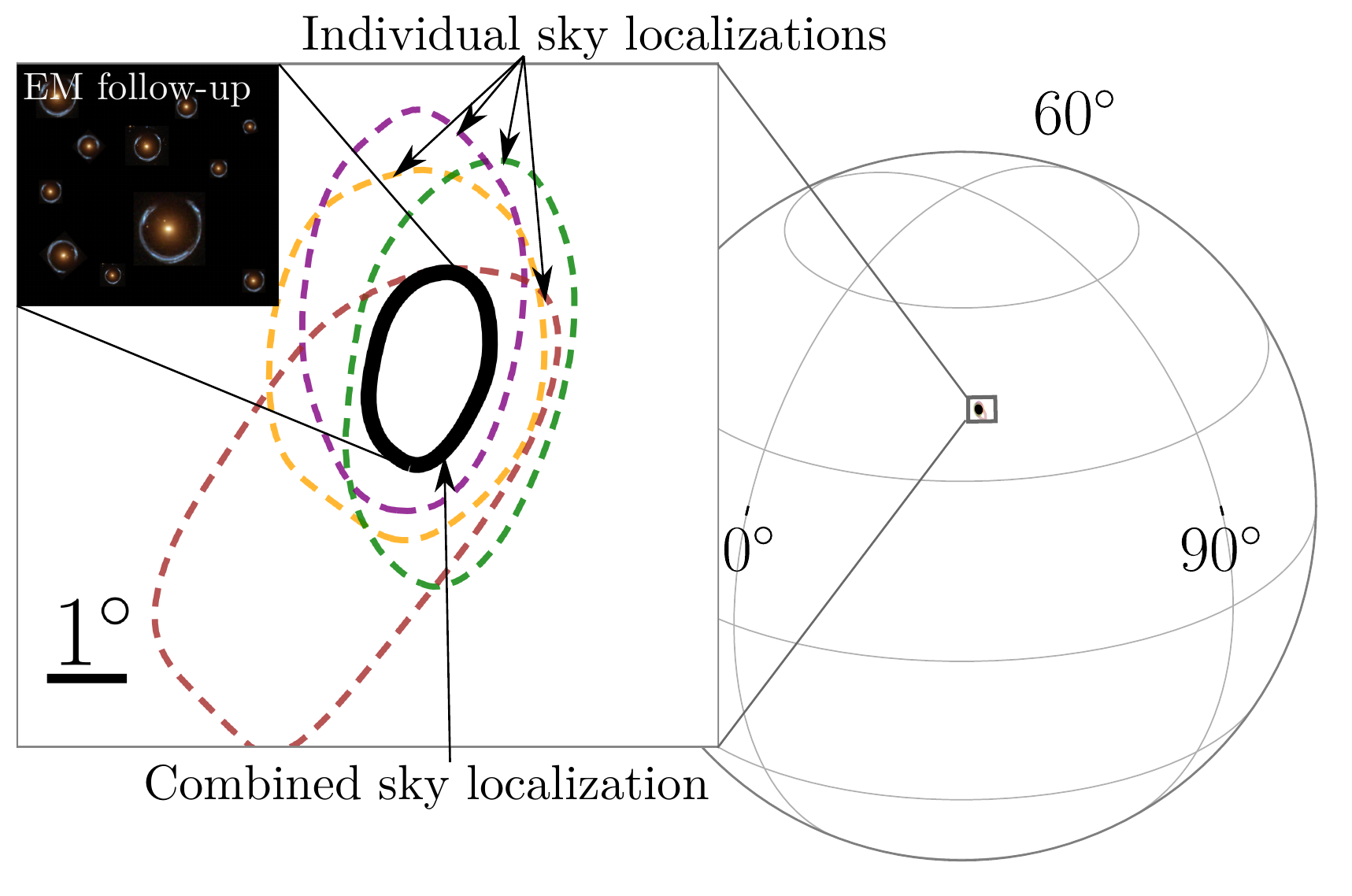}
    \caption{An illustration of a sky localization of a quadruply lensed gravitational wave. 
    We show both the individual (color) and the combined (black) sky localizations at 90\% confidence. 
    Each lensed gravitational-wave essentially gives us a new set of detectors with which to localize the event in the sky, allowing for improved sky localization. 
    A dedicated follow-up of the narrowed sky region would then allow us to search for the lensed host galaxy from which the gravitational-wave originates. 
    }
    \label{fig:skymapillustration}
\end{figure}

Moreover, because the Earth rotates during the delay between image arrivals, each image essentially gives us a new set of detectors \footnote{The typical image separation for strong lensing is less than an arcsecond \citep{collett2015}. Thus, the multiple images appear at essentially the same sky location given the GW detector accuracy.}. Due to the effectively larger detector network, strongly lensed events allow for better sky localization~\citep{Seto:2003iw}. 
Thus, a lensed event can be localized better in the sky than a non-lensed event (see Fig.~\ref{fig:skymapillustration} for illustration). 
However, even with the improved sky localization, there are still approximately 140 other lenses per square degree in the sky area \citep{collett2015}. 

In the case of doubly imaged GWs, there are two pieces of information immediately accessible to us from the GWs: the time delay between the images and the flux ratio of the images \citep{Sereno:2011ty,Yu:2020agu}. However, these two pieces of information alone will not significantly constrain the lensing system as the time delay is degenerate with the lens mass distribution, and the alignment of lens and source on the sky. 
Indeed, many of the lenses within the sky localization area will be consistent with a single time delay and magnification ratio\footnote{Although only massive cluster lenses can produce time delays of order years. The rarity of such clusters might mean that a pair of images with long time delay \textit{is} identifiable to a specific cluster lens, as investigated in Refs.~\citep{Smith:2017mqu,Smith:2018gle,smith2019discovery}}. 

Therefore, we limit our investigation to quadruple image systems. These systems have three independent time delay and magnification ratios: any lens system that cannot produce consistent time delays and magnification ratios cannot be the host of the lensed GW. Indeed, by combining the GW information with the information from the EM side, we can investigate if observations of a quadruply lensed GW event can provide a sufficiently unique fingerprint to definitively identify its host galaxy without an EM counterpart to the GW event.

Let us, therefore, make the following four assumptions: 
\begin{enumerate}
    \item We detect a quadruply imaged GW event.
    \item GW events originate within galaxies that emit EM radiation.
    \item We identify all of the strong lensing systems within the sky localization of the event.
    \item We have redshift information of each lens and source from EM observations.
\end{enumerate}
The first assumption is plausible when /Virgo reach design sensitivity: Single detector forecasts suggest $\sim 1$ strongly lensed event per year at LIGO design sensitivity \citep{2018MNRAS.476.2220L,PhysRevD.97.023012,2018MNRAS.480.3842O}. Moreover, \citep{2018MNRAS.476.2220L} found that $\sim 30\%$ of the detectable lensed events within LIGO would be quadruply lensed. In the third-generation detectors such as the Einstein Telescope~\citep{Maggiore:2019uih}, we could observe hundreds of lensed events~\citep{2014JCAP...10..080B,2015JCAP...12..006D}. These estimates assume that the signals that are below the noise threshold can not be detected. However, in the future, there exists an exciting possibility of identifying even some of the signals that are below the noise threshold~\cite{Li:2019osa,McIsaac:2019use}.

The second assumption should apply when the progenitors of binary black holes are stellar objects. BBH progenitors should trace the star formation rate or the stellar-mass, depending on the delay between massive black hole formation and BBH merger.  That the host galaxies emit EM radiation is widely applied in cosmography studies utilizing galaxy catalog based methods~\citep{Chen:2017rfc,Fishbach:2018gjp,Gray:2019ksv,Abbott:2019yzh,Soares-Santos:2019irc}.

The assumption that we know all of the lenses is challenging, even though we expect Euclid and LSST to find $\sim 10^5$ lenses \citep{collett2015}. Euclid lacks the depth to find every faint lensed source, and LSST lacks the angular resolution to detect small Einstein radius systems. However, there is no need to know the strong lenses at the moment the GW event is detected. If the sky localization is restricted to a few square degrees, then dedicated follow-up of this area with a wide field imaging space telescope like Euclid or WFIRST should quickly go deep enough to detect virtually all of the strongly lensed light (and hence stellar mass) originating at the typical redshifts of lensed GW events~\cite{Ryczanowski:2020mlt}.

Once the lenses are known, spectroscopic follow-up with a multi-object spectrograph (e.g., 4MOST, DESI, or Mauna Kea Spectroscopic Explorer) could be used to obtain redshifts for the lenses and sources. These facilities have thousands of spectroscopic fibers and fields-of-view of a few square degrees; hence they could simultaneously obtain all of the required redshifts in one or two very deep exposures.

\section{The catalog of strongly lensed binary black hole events}

Our simulated lens distribution follows the galaxy-galaxy lens population of ~\citep{collett2015}. The lenses are singular isothermal ellipsoid mass profiles with ellipticities and velocity dispersions following the observed distribution from SDSS \citep{choiparkvogeley}. We assume these potential lenses are uniformly distributed in a comoving volume out to $z=2$. Sources are then drawn from the Millennium Simulation \citep{springel2005} with galaxies painted on using a semi-analytic model \citep{delucia} and matched to the redshift distributions from the Hubble Ultra Deep Field \citep{connolley}. If the center of the source is multiply imaged, we include the system in our strong lens catalog. This catalog is complete down to sources with an $i$-band magnitude of 27. 

Our lensed GW population follows the lensed galaxy distribution: we treat every lensed source as equally likely to contain a lensed GW event (a more optimal method would involve luminosity and redshift weighting~\citep{Fan:2014kka,Chen:2017rfc,Fishbach:2018gjp,Gray:2019ksv,Abbott:2019yzh,Soares-Santos:2019irc}, but we leave this to future work). For the GW properties, we use a power-law black hole mass profile $p(m_1)\propto m^{-2.35}$ with a stellar-mass cut-off at $50 \, \rm M_\odot$ and uniform in mass ratio $q$, consistent with the LIGO/Virgo O1/O2 binary black hole population~\citep{LIGOScientific:2018jsj}. 
We use the \textsc{IMRPhenomPv2} waveform model~\cite{Hannam:2013oca,Husa:2015iqa,Khan:2015jqa}, which includes the full inspiral, ringdown, and merger parts of the GW signal, as implemented in the \textsc{LALSuite} software package~\citep{lalsuite}.  We infer the GW parameters using the \textsc{bilby} parameter inference software~\citep{2019ApJS..241...27A}.\footnote{Note that whilst \textsc{bilby} assumes non-lensed waveforms, this only affects the inferred luminosity distance and the phase of coalescence measurement. Other parameters are unbiased~\cite{1999PThPS.133..137N,Takahashi:2003ix,Dai:2017huk,2018arXiv180707062H,Pang:2020qow} }

A two or three detector network may have typical sky localization errors larger than we require here, and so we consider four gravitational-wave instruments. We assume the LIGO/Virgo/Kagra network at design sensitivity~\cite{TheLIGOScientific:2014jea,TheLIGOScientific:2016agk,TheVirgo:2014hva,Somiya:2011np,Aso:2013eba,Akutsu:2017kpk}, randomly simulate GWs that are quadruply lensed, and choose those that are detectable (i.e., all have a network signal-to-noise ratio $\rho_{\rm network}>10$). 

\section{Sky localization of multiply imaged events}

\begin{figure}
    \centering
    \includegraphics[width=\columnwidth]{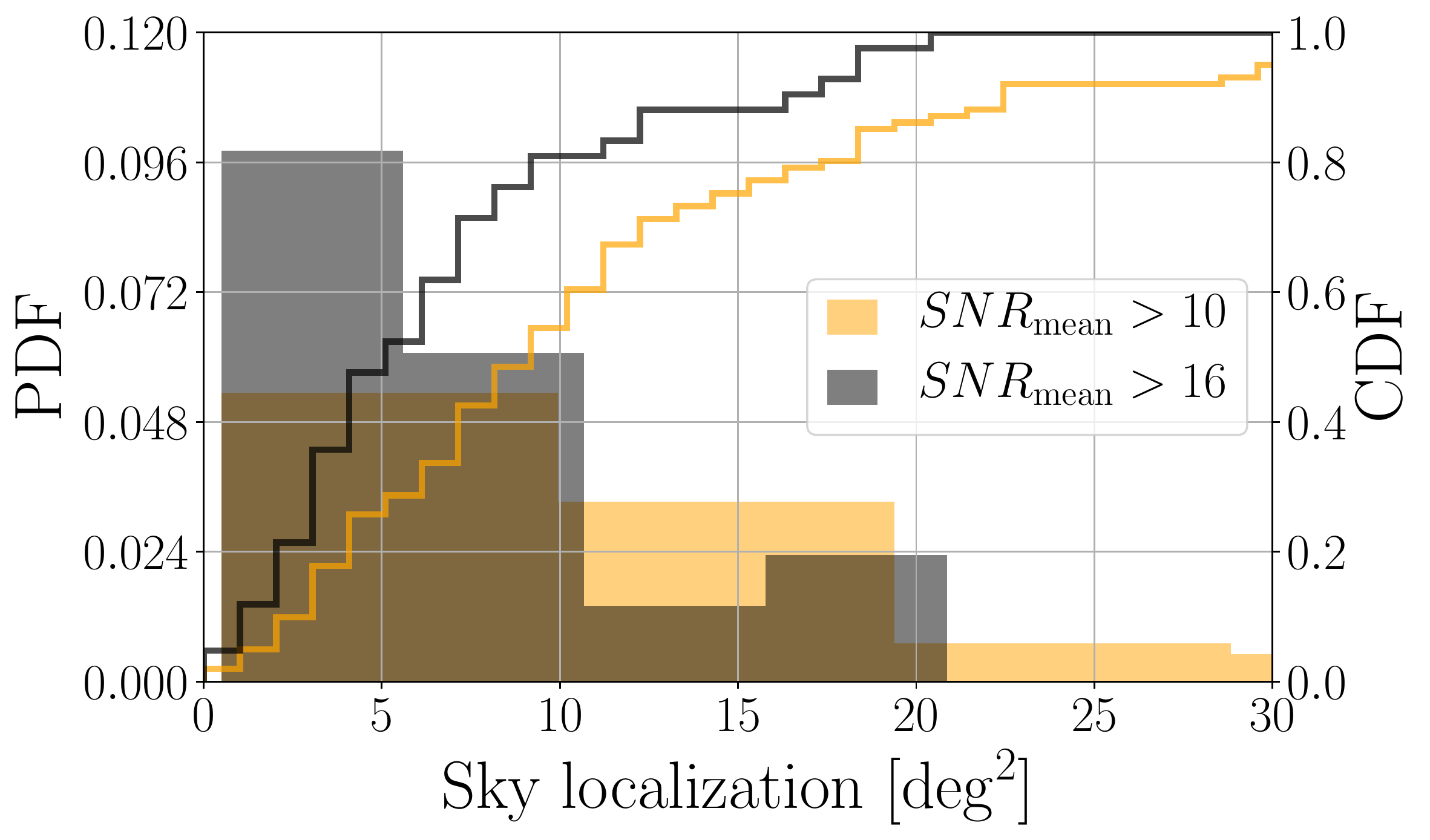}
    \caption{
    The probability (histogram) and cumulative distribution function (lines) of the combined sky localization constraints for our catalog of quadruply lensed events in the low (orange) and moderate (black) mean signal-to-noise ratio regimes. 
    We have combined the sky localization posteriors of the four individual lensed events. 
    At both low and moderate signal-to-noise ratio, a large fraction of the events are constrained to better than 10 $\deg^2$ in the sky, and often to better than $5$ $\deg^2$.
    We quote the $90\%$ confidence interval for the sky localization. 
    }
    \label{fig:gwskylocalization}
\end{figure}

We combine the sky localization posteriors of each image of the quadruply lensed GWs in our simulated catalog, finding that the typical sky localization of moderate (low) signal-to-noise ratio (SNR) detections is $< 10 \deg^2$ ($< 20 \deg^2$), and often a much lower $< 5 \deg^2$ ($<10 \deg^2$); see Fig.~\ref{fig:gwskylocalization}. Since we expect around $\sim 140$ lens galaxy candidates per square degree~\citep{collett2015}, quadruple image systems are immediately localized to $\sim\mathcal{O}(100)-\mathcal{O}(1000)$ host systems.

\section{Identifying the lens and source}
Once the event is localized, we can then ask the question 'which of the observed lenses can reproduce the observed time delays and magnifications?' Due to the computational costs of inverting the lens equation, we will not be able to perform our full search on a large statistical sample of detected GW injections. Instead, we choose three "representative" injections lensed by large (Einstein radius of $\sim 2 \, \rm arcsec$), medium ($\sim 1 \, \rm arcsec$) and small ($\sim 0.5 \, \rm arcsec$) lens. The simulated binary/lens systems are given in Tables~\ref{tab:injections1} and \ref{tab:injections2}. 
There are fewer massive lens systems, and they typically produce longer time delays. Thus, we expect that GW events with a longer time delay to be easier to identify.  Lower mass lenses are forecast to be more numerous \citep{collett2015}, so we expect that they will be harder to discriminate from each other. 

Within the sky localization of each event, we perform lens reconstruction of each possible lens to reproduce the observed time delays and magnification ratios. We model each lens as a singular power-law ellipsoidal mass distribution with external shear. The GW image positions are unknown, but we assume that we already know the lens model parameters to comparable precision to a rough initial lens model obtainable from ground-based imaging of the lensed EM host \citep{2014MNRAS.441.3238K}.
Specifically, we assume ($0.01$, $0.05$, $0.03$) spread (one-standard-deviation) on the measurement of the Einstein radius, axis-ratio, and each shear component, respectively. 
We assume the power-law density profile of the lens to be unconstrained by the existing data, adopting instead a prior typical of the strong lens population: a mean slope $\gamma=2$ with $0.2$ spread. These uncertainties are significantly broader than the errors achieved for detailed models of lenses with high-resolution imaging  \citep[e.g.][]{Birrer:2018vtm,collettauger2014,2019MNRAS.490.1743C}.
The errors also do not include the correlations between parameters whose inclusion would improve our discriminatory power and are thus conservative. To do the lens inversion, we use \textsc{lenstronomy}, a multi-purpose gravitational lens modeling software package~\citep{Birrer:2018xgm}. In the modeling, we neglect GW event timing uncertainty, but we add a 20 percent uncertainty on each image magnification to account for lensing by dark substructures.

We compute the Bayes factor for each lens within a sky localization of $4 \, \deg^2$ of the GW. Bayes factors are significant for lenses that can reproduce the observed lensed GW events, and low for lenses that are inconsistent with producing the observations. For detailed derivation, see the Methods section. 

\begin{figure}
    \centering
    \includegraphics[width=\columnwidth]{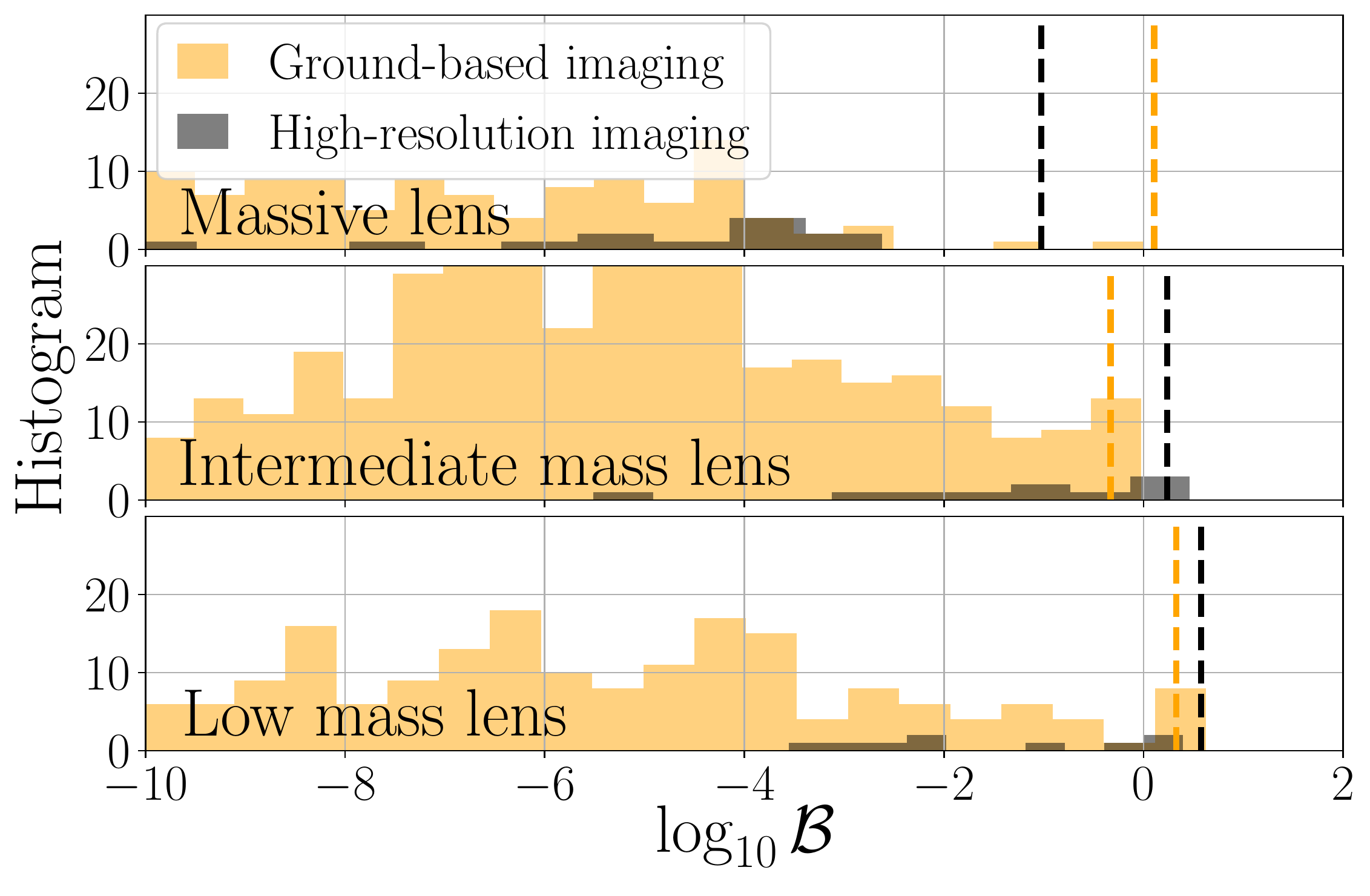}
    \caption{
    The Bayes factor in favor of a given galaxy being the host of the merging black hole.  
    We show the results for $550$ lens reconstructions within the sky localization of the injected gravitational-wave using ground-based imaging (orange). 
    The Bayes factor for the 11 best fitting lenses are shown in black after modeling simulated high-resolution follow-up imaging.  
    The Bayes factors are large for lenses that can reproduce the observed lensed GW event properties, and low for lenses that are inconsistent with producing the observations. 
    We show three lensing configurations: a gravitational-wave lensed by a massive $\sim 2 \, \rm arcsec$ Einstein radius lens (top panel), a moderate $\sim 1 \, \rm arcsec$ lens (middle panel), and a small $\sim 0.5\, \rm arcsec$ lens (bottom panel). The correct lens yields a high Bayes factor in all three cases (vertical dashed lines).  In the massive lens scenario (top panel), the background of lensed galaxy candidates is separated. 
    Thus, we can uniquely narrow down the source to one galaxy at above 90\% confidence with high-resolution imaging. In the moderate and small lens scenarios (middle and bottom panels), we narrow down the host galaxy to four and five candidates, respectively. 
    }
    \label{fig:gwselection}
\end{figure}

In our simulation, we find that the Bayes factor allows us to identify the host galaxy when the lens is massive enough (Fig.~\ref{fig:gwselection}, top panel, orange bins). For smaller lenses, we could narrow down the number of host galaxies to a few or a dozen (Fig.~\ref{fig:gwselection}, middle and bottom panels, orange bins). We can discriminate larger lenses more easily because they are rare, thus providing characteristic time-delay measurements that are produced by only a few similarly massive lenses. 

Systems with a few remaining candidates can be further narrowed down using detailed lens modeling. 
Therefore, based on the initial ground-based imaging results, we choose the 11 highest Bayes factor candidates\footnote{This could be the default analysis if automatic lens modeling \citep{Nightingale:2017cdh} can produce high fidelity lens models for every strong lens within the sky localization.} and model high-resolution imaging of each system. 
Specifically, we use \textsc{Lenstronomy} to reconstruct the lens properties observed with a simulated Hubble Space Telescope image. 
We then find that we can narrow down the lenses to one, four, and five for the large, medium, and small lens scenarios, respectively (Fig.~\ref{fig:gwselection}, black bins). 

Therefore, we can generally localize the GW source to one or at most a few galaxies. The number of potential host galaxy candidates scales with the sky area. Thus, moderate to high signal-to-noise ratio detections will be more promising and will allow us to discriminate the sources better. 
We expect more precise modeling of the lens and GW priors, and the inclusion of so-called joint-PE methodologies (Lo, private communication) to also improve our discriminatory power.

\section{Locating the BBH merger within the lensed host and measuring the Hubble constant}

\begin{figure}
    \centering
    \includegraphics[width=\columnwidth]{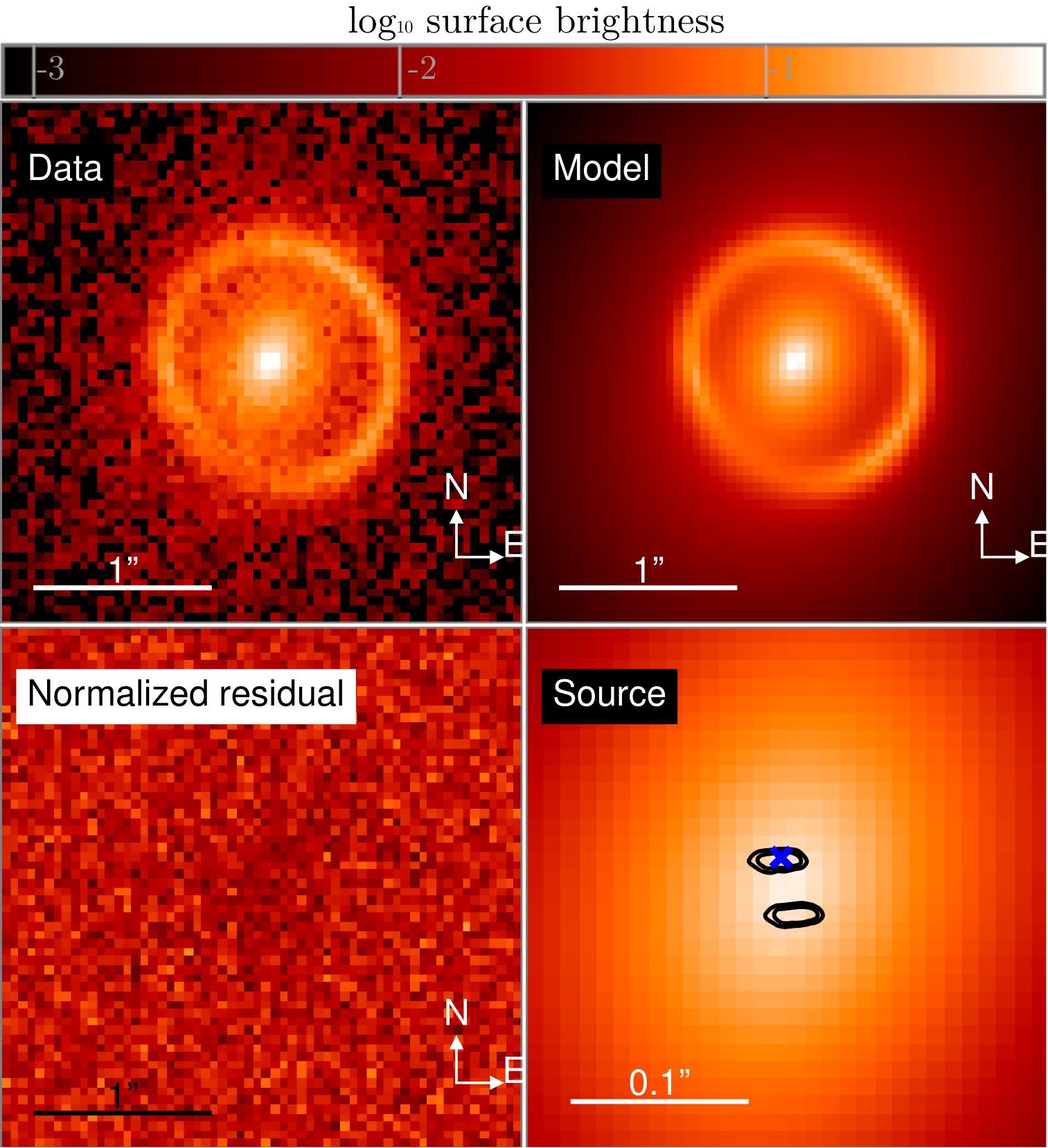}
    \caption{The sample lens system we use in our simulated Hubble constant measurement. Top-left panel: Observed light distribution. Top-right panel: Best fit model of the lens and the source. Bottom-left panel: The difference after subtracting the model from the data. Bottom-right panel: The reconstruction of the non-lensed source for the best fitting model, and the inferred position of the binary black hole relative to the source at 68\% and 90\% confidence (black contour) as well as its true position (blue cross).
    Our final sky localization is factor $\sim 10^9$ times better than the initial localization by LIGO/Virgo/Kagra ($\sim 4 \, \rm deg^2$).
    }
    \label{fig:lensreconstruction}
\end{figure}

Once the GW host system has been identified, a detailed lens model can be used to de-lens the EM source, identify which positions on the source plane can produce the observed time-delays and magnifications, and to convert time-delays and magnifications into inference on the Hubble constant. 
We use \textsc{Lenstronomy} to reconstruct a typical Einstein ring observed with a simulated Hubble Space Telescope image, shown in Fig.~\ref{fig:lensreconstruction}. 

We simulate random realizations of lensed GWs in this system until one of them is detected as a quadruple image event within LIGO/Virgo/Kagra. Given the lens model, the time delay ratios and magnification ratios localize the lens within the source. However, the symmetry of the lensing system means the source position is not uniquely determined. Marginalizing over the uncertainty in the lens and source parameters enables us to locate the BBH merger to one of two regions \footnote{The lens model localizes the source position to one of four regions, but these are blurred into two distinct regions after combining with the uncertainty on the source position inferred from the EM modeling}.

Since the ratio of time delays and magnifications is sufficient to constrain the source position, the absolute scale of the time delays and the absolute magnifications are still sensitive to the Hubble constant even without an EM counterpart. The time delays are sensitive to the Hubble constant through the time delay distance \citep{Refsdal1964,2017NatCo...8.1148L}, and the magnifications are sensitive through the luminosity distance to the GW source because BBHs can be regarded as standard sirens \citep{Schutz:1986gp,Abbott:2017xzu}. Converting the distances to cosmological parameter inference requires knowledge of lens and source redshifts, but these can be measured from in the EM for the lens and host.

\begin{figure}
    \centering
    \includegraphics[width=\columnwidth]{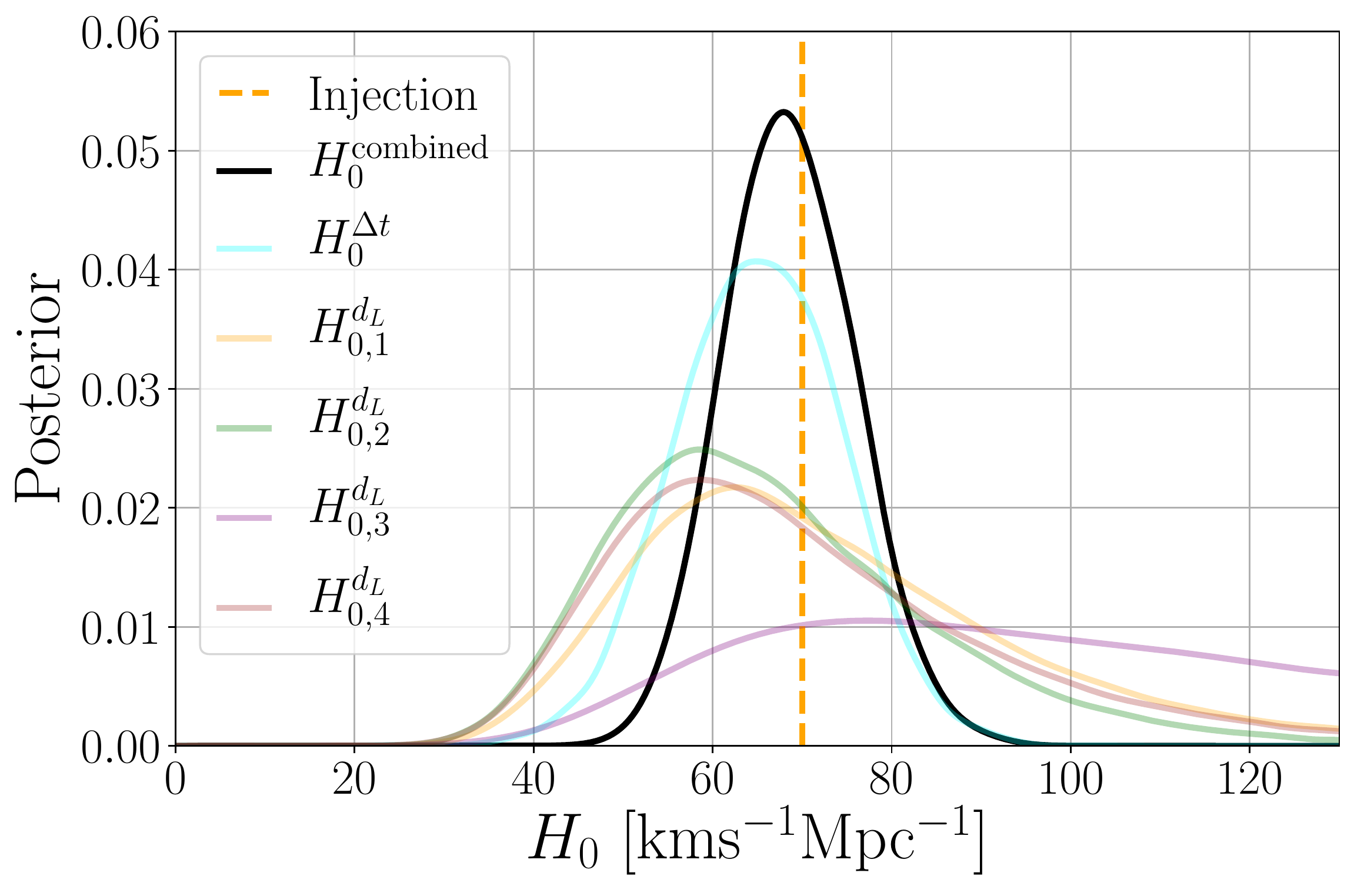}
    \caption{Measurement of the Hubble constant $H_0$ from the combination of gravitational-wave data and the lensed EM host galaxy  (black).  The coloured lines show the $H_0$ inferred from each of the four lensed gravitational wave 'standard siren' posteriors and the time-delay distance measurement (cyan). Combining these constraints yields a more stringent $H_0$ measurement compared to the individual posteriors. The dashed line shows the simulated value of $H_0=70 \, \rm km s^{-1} Mpc^{-1}$.
    }
    \label{fig:hubbleconstantmeasurement}
\end{figure}

To illustrate the cosmological sensitivity, we show constraints on the Hubble constant, in a flat $\Lambda$CDM cosmology with $\Omega_M$ fixed. 
We show the inferred $H_0$ in Fig.~\ref{fig:hubbleconstantmeasurement}. 
Combining the $H_0$ constraints from the time delay distance and the four images of the standard siren, we find ${H_0=68_{-6}^{+8}}$~km~s$^{-1}$~Mpc$^{-1}$ (median with symmetric 68\% credible interval). 

The primary contribution to the $H_0$ measurement comes from the measurement of the time-delay distance. 
The secondary contribution is from the standard siren. 
However, this is based on the assumption that there is a 20 percent scatter between the lens model magnification and the true magnification of the GW event. Such a discrepancy is possible, since the lens model is sensitive only to lensing structures with Einstein radii comparable to the size of the host, whereas the point GW source can also be (de-)magnified by stars and dark matter subhalos.\footnote{However, note that the error propagates only as $\sqrt{\mu}$ in the amplitude and thus the luminosity-distance measurement. Therefore, even significant magnification errors do not necessarily greatly propagate into the luminosity distance.} 20 percent scatter significantly smaller than expected for lensed supernovae \citep{Foxley-Marrable:2018dzu}, where stellar microlensing plays a significant role. However, microlensing is not expected to be as significant for lensed GWs as the Einstein radius of a star is smaller than the wavelength of stellar-mass BBH gravitational wave emission ~\citep{Oguri:2019fix} (except in the case of extreme macro model magnification~\citep{diego2019observational}). A dedicated study towards gravitational-wave microlensing will be needed to quantify a more realistic estimate of the magnification uncertainties due to microlensing. Such a study will require detailed wave optics modeling~\cite{Takahashi:2003ix} and is beyond the scope of this work.\footnote{See also studies on the possibility of wave optics effects affecting the time delay~\cite{Takahashi:2016jom,Cremonese:2018cyg,Morita:2019sau,Suyama:2020lbf,Ezquiaga:2020spg}}
However, even if we assume no information about the image magnifications, we find the constraint from the time-delay distance alone is $H_0=65 \pm 9$~km~s$^{-1}$~Mpc$^{-1}$.

For the Hubble constant inference, we have individually modeled each of the magnified standard sirens despite them being images of the same event; a joint-PE based parameter estimation could remove these excess degrees of freedom and improve the $H_0$ constraint (Lo, private communication).

\section{Discussion and Conclusion}

We have presented a method that can uniquely localize the host galaxies of gravitational waves from \emph{binary black holes} using quadruply imaged strongly lensed gravitational waves. The sky localization area of these sources will be $\sim$a billion times smaller than typical non-lensed BBHs. 

Binary black hole localization combined with an EM redshift measurement could open several new scientific frontiers. 
The Hubble constant and cosmological inference using lensed gravitational waves has been discussed in previous works in the context of events accompanied by an EM counterpart~\cite{2017NatCo...8.1148L,Cao:2019kgn}. We have $H_0$ measurements in the more common scenario of a quadruply imaged binary black hole without an EM counterpart. For higher redshift sources, these measurements also have some sensitivity to the density and equation of state of dark energy \citep{Sereno:2011ty}.\footnote{Cosmography with lensed GWs was also discussed in~\cite{Sun:2019ztn,Hou:2019dcm} in the context of beating patterns that occur when the time-delay is smaller than the signal length.} 
Other potential science cases include accurate tests of GW propagation and improved polarization tests with effectively more detectors~\citep[see Refs.][for discussion]{PhysRevD.86.022004,Smith:2018gle}.
and the interconnection between galaxies and black holes~\cite{LIGOScientific:2018jsj,Adhikari:2020wpn}. 

In some scenarios, we will only be able to localize the source host galaxy to a few candidates. These systems can still contribute to statistical studies. For example, we can perform cosmography studies by marginalizing the Hubble constant measurement according to the Bayes factor of each candidate. Comparable methods have been developed for Hubble constant measurements utilizing galaxy catalogs~\citep{Chen:2017rfc,Fishbach:2018gjp,Gray:2019ksv,Abbott:2019yzh,Soares-Santos:2019irc}. 

We note that our methodology can give an independent test of the lensed hypothesis if a host galaxy with consistent time-delay and magnification ratios is identified in a follow-up EM search. Three events or two time-delay ratios might be sufficient to perform the search~\citep{2020arXiv200712709D}. In the case of two events, a single time-delay estimate may be quite degenerate with the lens parameters and the source alignment, and it is not clear how well the search could be performed. We note that due to the rarity of galaxy clusters, the search is expected to be even more powerful for galaxy cluster lenses than for galaxy lenses. 

With the recently approved A+ detector upgrade~\citep{InstrumentationWhitePaper}, the sky localization should further improve.
As a consequence, lens identification becomes proportionately easier. In the third-generation detectors such as the Einstein Telescope~\citep{Maggiore:2019uih}, we could observe hundreds of lensed events at even higher signal-to-noise ratio~\citep{2014JCAP...10..080B,2015JCAP...12..006D}. 
In the era of future detectors identifying the hosts of quadruply imaged GWs should regularly be possible without an EM counterpart. 

\section{Acknowledgements}
We thank David Keitel, Ken Ng, Ignacio Magana Hernandez, Chris van den Broeck, Archisman Ghosh, Daniel Holz, Peter Pang, Rico Lo,  and Ian Harry for discussions, comments, and feedback. OAH is supported by the research program of the Netherlands Organization for Scientific Research (NWO). TEC is funded by a Royal Astronomical Society Research Fellowship and by a Dennis Sciama Fellowship from the University of Portsmouth. M\c{C} is supported by the Enrico Fermi Institute.
TGFL is partially supported by grants from the Research Grants Council of the Hong Kong (Project No. 14306218), Research Committee of the Chinese University of Hong Kong and the Croucher Foundation of Hong Kong.
The sky map in Fig.~\ref{fig:skymapillustration} uses the \textsc{ligo.skymap} package, and the lens image is a modified version of a Hubble Space Telescope image of the Cosmic Horseshoe. 
The authors are grateful for computational resources provided by the LIGO Laboratory and supported by National Science Foundation Grants PHY-0757058 and PHY-0823459. 

\section{Data availability}
The data underlying this article will be shared on reasonable request to the corresponding authors.

\newpage

\appendix

\section{Methodology}

\subsection{Determination of the Hubble constant from angular diameter distances}

For any given lens system, the Hubble constant is related to the time-delays measured from the GWs by~\citep{2017NatCo...8.1148L}
\begin{equation}
 \Delta t_{ij} = \frac{D_{\Delta t}(z,z_L,H_0^{\Delta t})(1+z_L)}{c} \Delta \phi_{ij}\,,
\end{equation}
where $z_s$ and $z_L$ are the source and lens redshifts, $\Delta \phi_{ij}$ is the reconstructed fermat potential at the image positions (for $i,j$ pairs of images),  $\Delta t_{ij}$ is the lensing time-delay between the two GW signals, and
\begin{equation} \label{eq:timedelaydistance}
    D_{\Delta t}(z,z_L,H_0^{\Delta t}) = \frac{d_{A}(z_L, H_0^{\Delta t}) d_{A}(z_s, H_0^{\Delta t})}{d_{A}(z_L,z_s, H_0^{\Delta t})}\,,
\end{equation}
is a combination of the angular diameter distances.

We can retrieve the fermat potential between the two images $\Delta \phi_{ij}$ and $D_{\Delta t}$ in \emph{unison} by solving the lens equation for a quad system. In particular, the lens system will have four source positions and unique time-delay distance for a given combination of GW time-delays $\{ t_i \}$. After solving the time-delay distance $D_{\Delta t}$, we can retrieve $H_0$ from Eq.~\ref{eq:timedelaydistance}. Its posterior distribution 
\begin{equation}
\begin{split}
    p(H_0^{\Delta t}|d_{\rm EM}^t) = &p(H_0^{\Delta t}|\vec{\theta}_L, z_L, z_s, \{ t_i \}) \\
    &p(\vec{\theta}_L,z_L, z_s, \{t_i\}|d_{\rm EM}^t) \,,
\end{split}
\end{equation}
where $d_{\rm EM}^t$ includes the EM data (lens reconstruction, redshift measurements) and the GW time-delay data. The posterior $p(\vec{\theta}_L,z_L, z_s, \{t_i\}|d_{\rm EM}^t)$ includes the lens parameters $\vec{\theta}_L$, the redshifts ($z_L, z_s$), and the GW time-delays $\{t_i\}$. 

\subsection{Determination of the magnification and luminosity distance from GWs} \label{sssec:gwproperties}

We can alternatively measure the Hubble constant by using the absolute image magnifications. To do so, we first need to match the relative magnification of the GW observations with those obtained from the lens reconstruction.

The GW measurement of the luminosity distance is fully degenerate with the magnification of the signal, i.e., 
\begin{equation}
\begin{split}
    D_{\rm obs}^i &= d_L / \sqrt{\mu_i}\,,
\end{split}
\end{equation}
where $D_{\rm obs}^i$ is the observed luminosity distance (as inferred from the GWs) of the $i$th signal and $\mu_i$ is the corresponding magnification. $d_L$ is the true luminosity distance of the object. 

Even without the complementary knowledge of the lens system, we can straightforwardly compute the relative magnification 
\begin{equation}
    \mu_r^{ij} = \frac{\mu_i}{\mu_j} = \left( \frac{D_{\rm obs}^j}{D_{\rm obs}^i} \right)^{2}\,.
\end{equation}
which is the division of the two \emph{observed} luminosity distance posteriors. The posterior (taking the dominant correlation between parameters to be between the inclination and luminosity distance) 
\begin{equation}
\begin{split}
    p(\mu_r^{ij} | d_{\rm GW}) \approx \int &  p(\mu_r^{ij} | D_{\rm obs}^i, D_{\rm obs}^j)  \\
    & \times \left( \frac{p(D_{\rm obs}^i, \iota| d_i)p(D_{\rm obs}^j, \iota| d_j)}{p(D_{\rm obs}^i|d_i) p(D_{\rm obs}^j|d_j) p(D_{\rm obs}^j) p(\iota)^2} \right) \\
    & \times p(\iota) p(D_{\rm obs}^i|d_i) p(D_{\rm obs}^j|d_j)
    d D_{\rm obs}^i d D_{\rm obs}^j d\iota \,, \\
\end{split}
\end{equation}
where $\iota$ is the inclination, which we included because it is highly degenerate with luminosity distance\footnote{If inclination was excluded, the relative magnification $\mu_r$ would be poorly constrained.}, and we assume a flat (agnostic) $\mu_{\rm rel}$ prior. The $d_{\rm GW}=\{d_i\}$ is the GW data strains, and $d_i$ is the $i$th observed GW strain. Note that in the lensing hypothesis all the parameters of the four signals will be the same, except for the observed luminosity-distance, time of coalescence, and phase of coalescence~\citep{2018arXiv180707062H}. Here we have assumed that the dominant correlation between luminosity distance and other parameters is the inclination, which is supported by several analyses detailing the luminosity distance-inclination degeneracy (e.g.~\citep{Mortlock:2018azx, Hotokezaka:2018dfi}). Nevertheless, including all correlations could slightly improve our measurement accuracies. 

\subsection{Determination of the lensing magnifications}
Given the lens model and the time-delays $\{t_i\}$, we obtain four potential source positions $\{\vec{y}^k\}$ with $k=0,1,2,3$ being the source position index (i.e., we obtain four source positions,  each of which will have four corresponding image positions). 
For each source position, we retrieve four lensing magnifications $\{\mu_i^k\}$, where $i$ is the image index and $k$ the source index. 
The posterior distribution 
\begin{equation}
\begin{split}
    p(\{\mu_i^k\}|d_{\rm EM}^t) = &p(\{\mu_i^k\}|\vec{\theta}_L, z_L, z_s, \{ t_i \}) \\
    & p(\vec{\theta}_L,z_L, z_s, \{t_i\}|d_{\rm EM}^t) d\vec{\theta}_L dz_L dz_s d\{t_i\} \,.
\end{split}
\end{equation}
The GWs give us the \emph{relative} magnifications $\mu_r^{ij} = \mu_i/\mu_j$ at a moderate accuracy. 
We can use this to test if the GW comes from this specific lensing system as follows:
Let the lensed and the null hypothesis be 
\begin{equation}
\begin{split}
\mathcal{H}_\mu: & \mu_r^{ij}+\delta \mu_r^{ij} = \widetilde{\mu}_r^{ij} = \mu_r^{ijk}=\frac{\mu_i^k}{\mu_j^k}\,\,\text{for some k}\,, \\
\mathcal{H}_0: & \mu_r^{ij} \, \, \text{and}\,\, \mu_r^{ijk}=\frac{\mu_i^k}{\mu_j^k}\,\,\text{are independent for all k}\,, \\
\end{split}
\end{equation}
where $\delta \mu_r^{ij}$ introduces a $20\%$ error spread due to milli- and micro-lensing. 
We've neglected the line of sight contribution to the H0 error, but this could be included in future work.
We note that microlensing can be suppressed for GWs due to \emph{diffraction} effects in the case of stellar-mass microlenses~\citep{Oguri:2019fix} (except in the case of extreme macromodel magnification~\citep{diego2019observational}). 
A dedicated study towards gravitational-wave microlensing will be needed to quantify a more realistic estimate of the magnification uncertainties due to microlensing. Such a study will require detailed wave optics modeling~\cite{Takahashi:2003ix} and is thus outside the scope of this work. 
The joint posterior $p(\widetilde{\mu}_r^{ij}, \mu_r^{ij}|d_{\rm GW},d_{\rm EM})$, contains the errors from gravitational-wave and lens parameter estimation, as well as the $20\%$ error spread due to milli- and micro-lensing.

The Bayes factor between the two hypotheses is then
\begin{equation}\label{eq:bayesfactor1}
\begin{split}
    \mathcal{M}^\mu_0 \approx \frac{1}{4} \sum_k \int \prod_{ij} & p(\widetilde{\mu}_r^{ij}|d_{\rm GW}, \mathcal{H}_0) p(\mu_r^{ij}|d_{\rm EM}^t, \mathcal{H}_0,k)\\
    & p(\mu_r^{ij}|\mathcal{H}_0)^{-1}d\mu_r^{ij}\,,
\end{split}
\end{equation}
where $p(\widetilde{\mu}_r^{ij}|d_{\rm GW}, \mathcal{H}_0)$ is the relative magnification from the GW luminosity distances only, marginalized over the microlensing error, $p(\mu_r^{ij}|d_{\rm EM}^t, \mathcal{H}_0,k)$ is the relative magnification predicted from the time-delay and the reconstructed lens for the source index $k$, and $p(\mu_r^{ij}|\mathcal{H}_0)$ is the relative magnification prior. The integral can be solved by importance sampling of the $p(\mu_r^{ij}|d_{\rm EM}^t, \mathcal{H}_0,k)$. We assume that the relative magnification prior is uniform; this assumption is roughly consistent with the findings in \citep{2018MNRAS.480.3842O}. Future studies are expected to assign more accurate priors as they become available. We stress that this will allow for a more \emph{optimal} definition of the Bayes factor but is not expected to hinder our analysis.

The correct source index $k$ is the one for which there is the largest evidence
\begin{equation} \label{eq:sourceindexevidence}
\begin{split}
    \frac{p(d|\mathcal{H}_\mu,k)}{p(d|\mathcal{H}_0)} = \prod_{ij}\int & p(\widetilde{\mu}_r^{ij}|\mathcal{H}_0, d_{\rm GW}) p(\mu_r^{ij}|d_{\rm EM}^t, \mathcal{H}_0,k)\\
    & p(\mu_r^{ij}|\mathcal{H}_0)^{-1}d\mu_r^{ij}\,,
\end{split}
\end{equation}
where $(i,j)$ run through $(0,1)$, $(1,2)$, $(2,3)$. We thus weight each sample by the evidence for the given source index. In principle, if the evidence for a given source index $k=k_0$ is substantial, we could retrieve the correct magnification posteriors
\begin{equation}
\begin{split}
    p(\{\mu_i\}|d_{\rm EM}^t) = &p(\{\mu_i\}|\vec{\theta}_L, z_L, z_s, \{ t_i \})
    p(\vec{\theta}_L,z_L, z_s, \{t_i\}|d_{\rm EM}^t)\,,
\end{split}
\end{equation}
where we have removed the index $k$ and assumed it to be $k=k_0$ to simplify the notation. However, note that we do not choose a specific source index in our analysis, or the one with the most substantial evidence. Instead, we (correctly) weigh each source index according to the evidence (Eq.~\ref{eq:sourceindexevidence}). 
In the scenarios that we have investigated, the most significant uncertainty in the magnification measurement comes from the GW measurement, followed by the milli/microlensing error, followed by the lens reconstruction error. Note that the magnification uncertainty propagates only as $\sqrt{\mu}$ in the amplitude and thus the luminosity-distance measurement. 

\subsection{Determination of the Hubble constant from luminosity distance}

After retrieving the image magnifications, we can estimate the Hubble constant a secondary way, using the host galaxy redshift $z_s$
\begin{equation}
    d_L(z_s,H_{0,i}^{d_L}) = \frac{(1+z_s) c}{H_{0,i}^{d_L}}F(z_s)\,.
\end{equation}
This allows us to measure the Hubble constant whose posterior 
\begin{equation}
\begin{split}
    p(H_{0,i}^{d_L}|d_{\rm GW}, d_{\rm EM}^t) = \int &p(H_{0,i}^{d_L}|z_s, d_L) p(z_s|d_{\rm EM}) \\
                          & p(d_L|d_{\rm GW}, \mu_i)\\
                          & p(\mu_i|d_{\rm EM}^t) d z_s d d_L d \mu_i\,,
\end{split}
\end{equation}
where we mark the $i$th image and the corresponding Hubble constant measurement $H_{0,i}^{d_L}$ with index $i$.

\subsection{Identifying the lens galaxy based on Hubble constant measurements}

Once we have measured the Hubble constants $H_0^{\{\Delta t, d_L\}}$, we can perform two additional tests to identify the correct lensed host galaxy. The Hubble constant $H_0^{\Delta t}$ must be within its expected prior range
\begin{equation}\label{eq:bayesfactor2}
    \mathcal{R}_0^\mu = \int \frac{p(H_0^{\Delta t}|d_{\rm EM}^t) p(H_0^{\Delta t}|\mathcal{H}_\mu^\prime)}{p(H_0^{\Delta t}|\mathcal{H}_0^\prime)} d H_0^{\Delta t}\,,
\end{equation}
where $p(H_0^{\Delta t}|\mathcal{H}_\mu^\prime)\in [60,80] \, \rm km s^{-1} Mpc^{-1}$ is the expected prior range, and $p(H_0^{\Delta t}|\mathcal{H}_0^\prime)$ is some much wider prior range corresponding to the case that the galaxy is not the host. Here we choose the wider prior to be  $H_0^{\Delta t} \in [0,1000] \, \rm km s^{-1} Mpc^{-1}$. In principle, we can retrieve a more accurate prior choice for $H_0^{\Delta t}$ by sampling the expected lens distribution. Doing so would likely improve discriminatory power. 

Likewise, the secondary Hubble constant measurement $H_{0,i}^{d_L}$ must be within the expected prior 
\begin{equation}\label{eq:bayesfactor3}
    \widetilde{\mathcal{R}}_0^\mu = \prod_i \int \frac{p(H_{0,i}^{d_L}|d_{\rm GW} , d_{\rm EM}^t) p(H_{0,i}^{d_L}|\mathcal{H}_\mu^\prime)}{p(H_{0,i}^{d_L}|\mathcal{H}_0^\prime)} d H_{0,i}^{d_L}\,.
\end{equation}
Therefore, the total log Bayes factor for/against the hypothesis that the GW originates from a given lens candidate is
\begin{equation} \label{eq:bayesfactortotal}
    \log \mathcal{B}^\mu_0 = \log \mathcal{M}^\mu_0 + \log  \mathcal{R}_0^\mu + \log \widetilde{\mathcal{R}}_0^\mu\,.
\end{equation}

\subsection{Combined sky localization of a lensed wave}

Given that we have detected a quadruply lensed gravitational wave, we can combine their sky localization posteriors simply by re-weighting:
\begin{equation}
 p({\rm ra}, {\rm dec}|d_1, d_2, d_3, d_4) \propto \frac{ \prod_{i=1}^{4} p({\rm ra}, {\rm dec}|d_i)}{p({\rm ra}, {\rm dec})^3}\,,
\end{equation}
where we neglect the correlations between the other gravitational-wave parameters and the sky localization, as well as selection effects. Their inclusion would improve our ability to localize the event in the sky.

\subsection{Localizing the merging black hole within the host galaxy and measuring the combined Hubble Constant}

Once the lensed host galaxy is identified, we can further localize the merging black hole within the galaxy. We can retrieve this source localization straightforwardly from the posterior of the source positions $p(\{\vec{y}^k\}|d_{\rm GW},d_{\rm EM}^t)$. 

To measure the final Hubble constant, we first combine the four individual luminosity distance measurements together (assuming uniform-in-comoving-volume prior and negligible correlations between luminosity distance other binary parameters) to retrieve $H_0^{d_L}$. We then combine $H_0^{d_L}$ with the Hubble constant from the time-delay distance $H_0^{\Delta t}$, assuming a flat $H_0$ prior. A more detailed modeling of the prior and the inclusion of selection effects could yield slightly more stringent results.

\begin{table}
\begin{tabular}{c c c c c c c c c c}
\hline
\hline
 & $m_1$ & $m_2$ & $z_L$ & $z_s$ & $\theta_E$ & $q$ & $\gamma$ & $\gamma_1$ & $\gamma_2$ \\
\hline
1 & $9 \, \rm M_\odot$ &  $7 \, \rm M_\odot$ & $0.17$ & $0.97$ & $2"$   & $0.9$ & $2.1$ & $0.04$   & $0.03$ \\
2 & $11 \, \rm M_\odot$ & $10 \, \rm M_\odot$& $0.16$ & $0.94$ & $1"$   & $0.8$ & $1.8$ & $0$      & $-0.02$ \\
3 & $7 \, \rm M_\odot$ &  $5 \, \rm M_\odot$ & $0.99$ & $1.60$ & $0.5"$ & $0.7$ & $1.7$ & $-0.01$  & $0.08$ \\
\hline
\hline
\end{tabular}
\caption{The binary masses $m_1$, $m_2$, the lens and source redshifts $z_L$ and $z_s$, Einstein radius $\theta_E$, axis ratio $q$, power-law slope $\gamma$ and the shears $\gamma_1$ and $\gamma_2$ of our simulated lensed signals.\label{tab:injections1}
}
\end{table}
\begin{table}
\begin{tabular}{c c c c c c c c}
\hline
\hline
 & $\Delta t^{12}$ & $\Delta t^{23}$ & $\Delta t^{34}$ & $\mu_1$ & $\mu_2$ & $\mu_3$ & $\mu_4$ \\
\hline
1 & $7.2$ days & $18.1$ days & $14.0$ days & $6.3$ & $7.1$   & $6.7$ & $5.0$ \\
2 & $4.6$ days & $3.3$ hours & $4.4$ hours & $5.0$ & $20.2$   & $12.9$ & $9.6$ \\
3 & $2.4$ days & $4.0$ days & $1.7$ days & $3.6$ & $4.2$ & $2.6$ & $2.0$ \\
\hline
\hline
\end{tabular}
\caption{Relative time-delays between signals $t^{ij}$ and the image magnifications $\{\mu_i\}$ for our simulated lensed signals.
\label{tab:injections2}
}
\end{table}

\bibliographystyle{mnras}

\begin{thebibliography}{}
\makeatletter
\relax
\def\mn@urlcharsother{\let\do\@makeother \do\$\do\&\do\#\do\^\do\_\do\%\do\~}
\def\mn@doi{\begingroup\mn@urlcharsother \@ifnextchar [ {\mn@doi@}
  {\mn@doi@[]}}
\def\mn@doi@[#1]#2{\def\@tempa{#1}\ifx\@tempa\@empty \href
  {http://dx.doi.org/#2} {doi:#2}\else \href {http://dx.doi.org/#2} {#1}\fi
  \endgroup}
\def\mn@eprint#1#2{\mn@eprint@#1:#2::\@nil}
\def\mn@eprint@arXiv#1{\href {http://arxiv.org/abs/#1} {{\tt arXiv:#1}}}
\def\mn@eprint@dblp#1{\href {http://dblp.uni-trier.de/rec/bibtex/#1.xml}
  {dblp:#1}}
\def\mn@eprint@#1:#2:#3:#4\@nil{\def\@tempa {#1}\def\@tempb {#2}\def\@tempc
  {#3}\ifx \@tempc \@empty \let \@tempc \@tempb \let \@tempb \@tempa \fi \ifx
  \@tempb \@empty \def\@tempb {arXiv}\fi \@ifundefined
  {mn@eprint@\@tempb}{\@tempb:\@tempc}{\expandafter \expandafter \csname
  mn@eprint@\@tempb\endcsname \expandafter{\@tempc}}}

\bibitem[\protect\citeauthoryear{Aasi et~al.}{Aasi
  et~al.}{2015}]{TheLIGOScientific:2014jea}
Aasi J.,  et~al., 2015, \mn@doi [Class. Quant. Grav.]
  {10.1088/0264-9381/32/7/074001}, 32, 074001

\bibitem[\protect\citeauthoryear{Abbott et~al.}{Abbott
  et~al.}{2016}]{TheLIGOScientific:2016agk}
Abbott B.~P.,  et~al., 2016, \mn@doi [Phys. Rev. Lett.]
  {10.1103/PhysRevLett.116.131103}, 116, 131103

\bibitem[\protect\citeauthoryear{Abbott et~al.}{Abbott
  et~al.}{2017a}]{TheLIGOScientific:2017qsa}
Abbott B.~P.,  et~al., 2017a, \mn@doi [Phys. Rev. Lett.]
  {10.1103/PhysRevLett.119.161101}, 119, 161101

\bibitem[\protect\citeauthoryear{Abbott et~al.}{Abbott
  et~al.}{2017b}]{Abbott:2017xzu}
Abbott B.~P.,  et~al., 2017b, \mn@doi [Nature] {10.1038/nature24471}, 551, 85

\bibitem[\protect\citeauthoryear{Abbott et~al.}{Abbott
  et~al.}{2018}]{InstrumentationWhitePaper}
Abbott B.~P.,  et~al., 2018, DCC entry, LIGO-T1800133

\bibitem[\protect\citeauthoryear{Abbott et~al.}{Abbott
  et~al.}{2019a}]{Abbott:2019yzh}
Abbott B.,  et~al., 2019a, arXiv e-prints, \href
  {https://ui.adsabs.harvard.edu/abs/2019arXiv190806060T} {p. arXiv:1908.06060}

\bibitem[\protect\citeauthoryear{Abbott et~al.}{Abbott
  et~al.}{2019b}]{LIGOScientific:2018mvr}
Abbott B.,  et~al., 2019b, \mn@doi [Phys. Rev. X] {10.1103/PhysRevX.9.031040},
  9, 031040

\bibitem[\protect\citeauthoryear{Abbott et~al.}{Abbott
  et~al.}{2019c}]{LIGOScientific:2019fpa}
Abbott B.,  et~al., 2019c, \mn@doi [Phys. Rev. D]
  {10.1103/PhysRevD.100.104036}, 100, 104036

\bibitem[\protect\citeauthoryear{Abbott et~al.}{Abbott
  et~al.}{2019d}]{LIGOScientific:2018jsj}
Abbott B.~P.,  et~al., 2019d, \mn@doi [Astrophys. J.]
  {10.3847/2041-8213/ab3800}, 882, L24

\bibitem[\protect\citeauthoryear{Abbott et~al.}{Abbott
  et~al.}{2020}]{LIGOScientific:2020stg}
Abbott R.,  et~al., 2020, arXiv e-prints, \href
  {https://ui.adsabs.harvard.edu/abs/2020arXiv200408342T} {p. arXiv:2004.08342}

\bibitem[\protect\citeauthoryear{Acernese et~al.}{Acernese
  et~al.}{2015}]{TheVirgo:2014hva}
Acernese F.,  et~al., 2015, \mn@doi [Class. Quant. Grav.]
  {10.1088/0264-9381/32/2/024001}, 32, 024001

\bibitem[\protect\citeauthoryear{Adhikari, Fishbach, Holz, Wechsler  \&
  Fang}{Adhikari et~al.}{2020}]{Adhikari:2020wpn}
Adhikari S.,  Fishbach M.,  Holz D.~E.,  Wechsler R.~H.,   Fang Z.,  2020,
  arXiv preprint arXiv:2001.01025

\bibitem[\protect\citeauthoryear{Akutsu et~al.}{Akutsu
  et~al.}{2018}]{Akutsu:2017kpk}
Akutsu T.,  et~al., 2018, \mn@doi [PTEP] {10.1093/ptep/ptx180}, 2018, 013F01

\bibitem[\protect\citeauthoryear{{Ashton} et~al.,}{{Ashton}
  et~al.}{2019}]{2019ApJS..241...27A}
{Ashton} G.,  et~al., 2019, \mn@doi [\apjs] {10.3847/1538-4365/ab06fc}, \href
  {https://ui.adsabs.harvard.edu/abs/2019ApJS..241...27A} {241, 27}

\bibitem[\protect\citeauthoryear{Aso, Michimura, Somiya, Ando, Miyakawa,
  Sekiguchi, Tatsumi  \& Yamamoto}{Aso et~al.}{2013}]{Aso:2013eba}
Aso Y.,  Michimura Y.,  Somiya K.,  Ando M.,  Miyakawa O.,  Sekiguchi T.,
  Tatsumi D.,   Yamamoto H.,  2013, \mn@doi [Phys. Rev.]
  {10.1103/PhysRevD.88.043007}, D88, 043007

\bibitem[\protect\citeauthoryear{Berti et~al.}{Berti
  et~al.}{2015}]{Berti:2015itd}
Berti E.,  et~al., 2015, \mn@doi [Class. Quant. Grav.]
  {10.1088/0264-9381/32/24/243001}, 32, 243001

\bibitem[\protect\citeauthoryear{{Biesiada}, {Ding}, {Pi{\'o}rkowska}  \&
  {Zhu}}{{Biesiada} et~al.}{2014}]{2014JCAP...10..080B}
{Biesiada} M.,  {Ding} X.,  {Pi{\'o}rkowska} A.,   {Zhu} Z.-H.,  2014, \mn@doi
  [\jcap] {10.1088/1475-7516/2014/10/080}, \href
  {https://ui.adsabs.harvard.edu/abs/2014JCAP...10..080B} {2014, 080}

\bibitem[\protect\citeauthoryear{Birrer \& Amara}{Birrer \&
  Amara}{2018}]{Birrer:2018xgm}
Birrer S.,  Amara A.,  2018, Physics of the Dark Universe, 22, 189

\bibitem[\protect\citeauthoryear{Birrer et~al.}{Birrer
  et~al.}{2019}]{Birrer:2018vtm}
Birrer S.,  et~al., 2019, \mn@doi [Mon. Not. Roy. Astron. Soc.]
  {10.1093/mnras/stz200}, 484, 4726

\bibitem[\protect\citeauthoryear{Calderón~Bustillo, Laguna  \&
  Shoemaker}{Calderón~Bustillo et~al.}{2017}]{Bustillo:2016gid}
Calderón~Bustillo J.,  Laguna P.,   Shoemaker D.,  2017, \mn@doi [Phys. Rev.]
  {10.1103/PhysRevD.95.104038}, D95, 104038

\bibitem[\protect\citeauthoryear{Cao, Qi, Cao, Biesiada, Li, Pan  \& Zhu}{Cao
  et~al.}{2019}]{Cao:2019kgn}
Cao S.,  Qi J.,  Cao Z.,  Biesiada M.,  Li J.,  Pan Y.,   Zhu Z.-H.,  2019,
  \mn@doi [Sci. Rep.] {10.1038/s41598-019-47616-4}, 9, 11608

\bibitem[\protect\citeauthoryear{Chatziioannou, Yunes  \&
  Cornish}{Chatziioannou et~al.}{2012}]{PhysRevD.86.022004}
Chatziioannou K.,  Yunes N.,   Cornish N.,  2012, \mn@doi [Phys. Rev. D]
  {10.1103/PhysRevD.86.022004}, 86, 022004

\bibitem[\protect\citeauthoryear{Chatziioannou et~al.}{Chatziioannou
  et~al.}{2019}]{Chatziioannou:2019dsz}
Chatziioannou K.,  et~al., 2019, \mn@doi [Phys. Rev.]
  {10.1103/PhysRevD.100.104015}, D100, 104015

\bibitem[\protect\citeauthoryear{Chen \& Holz}{Chen \&
  Holz}{2016}]{Chen:2016tys}
Chen H.-Y.,  Holz D.~E.,  2016, arXiv preprint arXiv:1612.01471

\bibitem[\protect\citeauthoryear{Chen, Fishbach  \& Holz}{Chen
  et~al.}{2018}]{Chen:2017rfc}
Chen H.-Y.,  Fishbach M.,   Holz D.~E.,  2018, \mn@doi [Nature]
  {10.1038/s41586-018-0606-0}, 562, 545

\bibitem[\protect\citeauthoryear{{Chen} et~al.,}{{Chen}
  et~al.}{2019}]{2019MNRAS.490.1743C}
{Chen} G. C.~F.,  et~al., 2019, \mn@doi [\mnras] {10.1093/mnras/stz2547}, \href
  {https://ui.adsabs.harvard.edu/abs/2019MNRAS.490.1743C} {490, 1743}

\bibitem[\protect\citeauthoryear{{Choi}, {Park}  \& {Vogeley}}{{Choi}
  et~al.}{2007}]{choiparkvogeley}
{Choi} Y.-Y.,  {Park} C.,   {Vogeley} M.~S.,  2007, \mn@doi [\apj]
  {10.1086/511060}, \href
  {https://ui.adsabs.harvard.edu/abs/2007ApJ...658..884C} {658, 884}

\bibitem[\protect\citeauthoryear{Coleman~Miller \& Yunes}{Coleman~Miller \&
  Yunes}{2019}]{Miller:2019vfc}
Coleman~Miller M.,  Yunes N.,  2019, \mn@doi [Nature]
  {10.1038/s41586-019-1129-z}, 568, 469

\bibitem[\protect\citeauthoryear{{Collett}}{{Collett}}{2015}]{collett2015}
{Collett} T.~E.,  2015, \mn@doi [\apj] {10.1088/0004-637X/811/1/20}, \href
  {https://ui.adsabs.harvard.edu/abs/2015ApJ...811...20C} {811, 20}

\bibitem[\protect\citeauthoryear{{Collett} \& {Auger}}{{Collett} \&
  {Auger}}{2014}]{collettauger2014}
{Collett} T.~E.,  {Auger} M.~W.,  2014, \mn@doi [\mnras]
  {10.1093/mnras/stu1190}, \href
  {https://ui.adsabs.harvard.edu/abs/2014MNRAS.443..969C} {443, 969}

\bibitem[\protect\citeauthoryear{{Collett} \& {Bacon}}{{Collett} \&
  {Bacon}}{2016}]{collettbacon2016}
{Collett} T.~E.,  {Bacon} D.~J.,  2016, \mn@doi [\mnras]
  {10.1093/mnras/stv2791}, \href
  {https://ui.adsabs.harvard.edu/abs/2016MNRAS.456.2210C} {456, 2210}

\bibitem[\protect\citeauthoryear{Collett et~al.}{Collett
  et~al.}{2017}]{Collett:2017ksf}
Collett T.~E.,  et~al., 2017, \mn@doi [Astrophys. J.]
  {10.3847/1538-4357/aa76e6}, 843, 148

\bibitem[\protect\citeauthoryear{{Connolly} et~al.,}{{Connolly}
  et~al.}{2010}]{connolley}
{Connolly} A.~J.,  et~al., 2010, {Simulating the LSST system}.
p. 77381O, \mn@doi{10.1117/12.857819}

\bibitem[\protect\citeauthoryear{Cremonese \& Mörtsell}{Cremonese \&
  Mörtsell}{2018}]{Cremonese:2018cyg}
Cremonese P.,  Mörtsell E.,  2018, arXiv preprint arXiv:1808.05886

\bibitem[\protect\citeauthoryear{Dahle et~al.}{Dahle
  et~al.}{2013}]{Dahle:2012bd}
Dahle H.,  et~al., 2013, \mn@doi [Astrophys. J.] {10.1088/0004-637X/773/2/146},
  773, 146

\bibitem[\protect\citeauthoryear{Dai \& Venumadhav}{Dai \&
  Venumadhav}{2017}]{Dai:2017huk}
Dai L.,  Venumadhav T.,  2017, arXiv preprint arXiv:1702.04724

\bibitem[\protect\citeauthoryear{{Dai}, {Zackay}, {Venumadhav}, {Roulet}  \&
  {Zaldarriaga}}{{Dai} et~al.}{2020}]{2020arXiv200712709D}
{Dai} L.,  {Zackay} B.,  {Venumadhav} T.,  {Roulet} J.,   {Zaldarriaga} M.,
  2020, arXiv e-prints, \href
  {https://ui.adsabs.harvard.edu/abs/2020arXiv200712709D} {p. arXiv:2007.12709}

\bibitem[\protect\citeauthoryear{{De Lucia} \& {Blaizot}}{{De Lucia} \&
  {Blaizot}}{2007}]{delucia}
{De Lucia} G.,  {Blaizot} J.,  2007, \mn@doi [\mnras]
  {10.1111/j.1365-2966.2006.11287.x}, \href
  {https://ui.adsabs.harvard.edu/abs/2007MNRAS.375....2D} {375, 2}

\bibitem[\protect\citeauthoryear{{Diego, J. M.}, {Hannuksela, O. A.}, {Kelly,
  P. L.}, {Pagano, G.}, {Broadhurst, T.}, {Kim, K.}, {Li, T. G. F.}  \& {Smoot,
  G. F.}}{{Diego, J. M.} et~al.}{2019}]{diego2019observational}
{Diego, J. M.} {Hannuksela, O. A.} {Kelly, P. L.} {Pagano, G.} {Broadhurst, T.}
  {Kim, K.} {Li, T. G. F.}  {Smoot, G. F.} 2019, \mn@doi [A\&A]
  {10.1051/0004-6361/201935490}, 627, A130

\bibitem[\protect\citeauthoryear{{Ding}, {Biesiada}  \& {Zhu}}{{Ding}
  et~al.}{2015}]{2015JCAP...12..006D}
{Ding} X.,  {Biesiada} M.,   {Zhu} Z.-H.,  2015, \mn@doi [\jcap]
  {10.1088/1475-7516/2015/12/006}, \href
  {https://ui.adsabs.harvard.edu/abs/2015JCAP...12..006D} {2015, 006}

\bibitem[\protect\citeauthoryear{Ezquiaga, Hu  \& Lagos}{Ezquiaga
  et~al.}{2020}]{Ezquiaga:2020spg}
Ezquiaga J.~M.,  Hu W.,   Lagos M.,  2020, \mn@doi [Phys. Rev. D]
  {10.1103/PhysRevD.102.023531}, 102, 023531

\bibitem[\protect\citeauthoryear{Fan, Messenger  \& Heng}{Fan
  et~al.}{2014}]{Fan:2014kka}
Fan X.,  Messenger C.,   Heng I.~S.,  2014, \mn@doi [Astrophys. J.]
  {10.1088/0004-637X/795/1/43}, 795, 43

\bibitem[\protect\citeauthoryear{Fishbach et~al.}{Fishbach
  et~al.}{2019}]{Fishbach:2018gjp}
Fishbach M.,  et~al., 2019, \mn@doi [Astrophys. J.] {10.3847/2041-8213/aaf96e},
  871, L13

\bibitem[\protect\citeauthoryear{Foxley-Marrable, Collett, Vernardos, Goldstein
   \& Bacon}{Foxley-Marrable et~al.}{2018}]{Foxley-Marrable:2018dzu}
Foxley-Marrable M.,  Collett T.~E.,  Vernardos G.,  Goldstein D.~A.,   Bacon
  D.,  2018, \mn@doi [Mon. Not. Roy. Astron. Soc.] {10.1093/mnras/sty1346},
  478, 5081

\bibitem[\protect\citeauthoryear{Gray et~al.,}{Gray
  et~al.}{2020}]{Gray:2019ksv}
Gray R.,  et~al., 2020, Physical Review D, 101, 122001

\bibitem[\protect\citeauthoryear{Hannam, Schmidt, Bohé, Haegel, Husa, Ohme,
  Pratten  \& Pürrer}{Hannam et~al.}{2014}]{Hannam:2013oca}
Hannam M.,  Schmidt P.,  Bohé A.,  Haegel L.,  Husa S.,  Ohme F.,  Pratten G.,
    Pürrer M.,  2014, \mn@doi [Phys. Rev. Lett.]
  {10.1103/PhysRevLett.113.151101}, 113, 151101

\bibitem[\protect\citeauthoryear{Hannuksela, Haris, Ng, Kumar, Mehta, Keitel,
  Li  \& Ajith}{Hannuksela et~al.}{2019}]{Hannuksela_2019}
Hannuksela O.~A.,  Haris K.,  Ng K. K.~Y.,  Kumar S.,  Mehta A.~K.,  Keitel D.,
   Li T. G.~F.,   Ajith P.,  2019, \mn@doi [The Astrophysical Journal]
  {10.3847/2041-8213/ab0c0f}, 874, L2

\bibitem[\protect\citeauthoryear{{Haris}, {Mehta}, {Kumar}, {Venumadhav}  \&
  {Ajith}}{{Haris} et~al.}{2018}]{2018arXiv180707062H}
{Haris} K.,  {Mehta} A.~K.,  {Kumar} S.,  {Venumadhav} T.,   {Ajith} P.,  2018,
  arXiv e-prints, \href {https://ui.adsabs.harvard.edu/abs/2018arXiv180707062H}
  {p. arXiv:1807.07062}

\bibitem[\protect\citeauthoryear{Hotokezaka, Nakar, Gottlieb, Nissanke, Masuda,
  Hallinan, Mooley  \& Deller}{Hotokezaka et~al.}{2019}]{Hotokezaka:2018dfi}
Hotokezaka K.,  Nakar E.,  Gottlieb O.,  Nissanke S.,  Masuda K.,  Hallinan G.,
   Mooley K.~P.,   Deller A.,  2019, \mn@doi [Nat. Astron.]
  {10.1038/s41550-019-0820-1}, 3, 940

\bibitem[\protect\citeauthoryear{Hou, Fan, Liao  \& Zhu}{Hou
  et~al.}{2020}]{Hou:2019dcm}
Hou S.,  Fan X.-L.,  Liao K.,   Zhu Z.-H.,  2020, \mn@doi [Phys. Rev. D]
  {10.1103/PhysRevD.101.064011}, 101, 064011

\bibitem[\protect\citeauthoryear{Husa, Khan, Hannam, Pürrer, Ohme,
  Jiménez~Forteza  \& Bohé}{Husa et~al.}{2016}]{Husa:2015iqa}
Husa S.,  Khan S.,  Hannam M.,  Pürrer M.,  Ohme F.,  Jiménez~Forteza X.,
  Bohé A.,  2016, \mn@doi [Phys. Rev.] {10.1103/PhysRevD.93.044006}, D93,
  044006

\bibitem[\protect\citeauthoryear{Khan, Husa, Hannam, Ohme, Pürrer,
  Jiménez~Forteza  \& Bohé}{Khan et~al.}{2016}]{Khan:2015jqa}
Khan S.,  Husa S.,  Hannam M.,  Ohme F.,  Pürrer M.,  Jiménez~Forteza X.,
  Bohé A.,  2016, \mn@doi [Phys. Rev.] {10.1103/PhysRevD.93.044007}, D93,
  044007

\bibitem[\protect\citeauthoryear{{Kostrzewa-Rutkowska}, {Wyrzykowski}, {Auger},
  {Collett}  \& {Belokurov}}{{Kostrzewa-Rutkowska}
  et~al.}{2014}]{2014MNRAS.441.3238K}
{Kostrzewa-Rutkowska} Z.,  {Wyrzykowski} {\L}.,  {Auger} M.~W.,  {Collett}
  T.~E.,   {Belokurov} V.,  2014, \mn@doi [\mnras] {10.1093/mnras/stu783},
  \href {https://ui.adsabs.harvard.edu/abs/2014MNRAS.441.3238K} {441, 3238}

\bibitem[\protect\citeauthoryear{{LIGO Scientific Collaboration}}{{LIGO
  Scientific Collaboration}}{2018}]{lalsuite}
{LIGO Scientific Collaboration} 2018, {LIGO} {A}lgorithm {L}ibrary -
  {LALS}uite, free software (GPL), \mn@doi{10.7935/GT1W-FZ16}

\bibitem[\protect\citeauthoryear{{Li}, {Mao}, {Zhao}  \& {Lu}}{{Li}
  et~al.}{2018}]{2018MNRAS.476.2220L}
{Li} S.-S.,  {Mao} S.,  {Zhao} Y.,   {Lu} Y.,  2018, \mn@doi [\mnras]
  {10.1093/mnras/sty411}, \href
  {https://ui.adsabs.harvard.edu/abs/2018MNRAS.476.2220L} {476, 2220}

\bibitem[\protect\citeauthoryear{Li, Lo, Sachdev, Chan, Lin, Li  \&
  Weinstein}{Li et~al.}{2019}]{Li:2019osa}
Li A. K.~Y.,  Lo R. K.~L.,  Sachdev S.,  Chan C.~L.,  Lin E.~T.,  Li T. G.~F.,
   Weinstein A.~J.,  2019, arXiv preprint arXiv:1904.06020

\bibitem[\protect\citeauthoryear{{Liao}, {Fan}, {Ding}, {Biesiada}  \&
  {Zhu}}{{Liao} et~al.}{2017}]{2017NatCo...8.1148L}
{Liao} K.,  {Fan} X.-L.,  {Ding} X.,  {Biesiada} M.,   {Zhu} Z.-H.,  2017,
  \mn@doi [Nature Communications] {10.1038/s41467-017-01152-9}, \href
  {https://ui.adsabs.harvard.edu/abs/2017NatCo...8.1148L} {8, 1148}

\bibitem[\protect\citeauthoryear{Maggiore et~al.}{Maggiore
  et~al.}{2020}]{Maggiore:2019uih}
Maggiore M.,  et~al., 2020, Journal of Cosmology and Astroparticle Physics,
  2020, 050

\bibitem[\protect\citeauthoryear{McIsaac, Keitel, Collett, Harry, Mozzon, Edy
  \& Bacon}{McIsaac et~al.}{2019}]{McIsaac:2019use}
McIsaac C.,  Keitel D.,  Collett T.,  Harry I.,  Mozzon S.,  Edy O.,   Bacon
  D.,  2019, arXiv e-prints, \href
  {https://ui.adsabs.harvard.edu/abs/2019arXiv191205389M} {p. arXiv:1912.05389}

\bibitem[\protect\citeauthoryear{Morita \& Soda}{Morita \&
  Soda}{2019}]{Morita:2019sau}
Morita T.,  Soda J.,  2019, arXiv preprint arXiv:1911.07435

\bibitem[\protect\citeauthoryear{Mortlock, Feeney, Peiris, Williamson  \&
  Nissanke}{Mortlock et~al.}{2019}]{Mortlock:2018azx}
Mortlock D.~J.,  Feeney S.~M.,  Peiris H.~V.,  Williamson A.~R.,   Nissanke
  S.~M.,  2019, \mn@doi [Phys. Rev.] {10.1103/PhysRevD.100.103523}, D100,
  103523

\bibitem[\protect\citeauthoryear{{Nakamura} \& {Deguchi}}{{Nakamura} \&
  {Deguchi}}{1999}]{1999PThPS.133..137N}
{Nakamura} T.~T.,  {Deguchi} S.,  1999, \mn@doi [Progress of Theoretical
  Physics Supplement] {10.1143/PTPS.133.137}, \href
  {https://ui.adsabs.harvard.edu/abs/1999PThPS.133..137N} {133, 137}

\bibitem[\protect\citeauthoryear{Ng, Wong, Broadhurst  \& Li}{Ng
  et~al.}{2018}]{PhysRevD.97.023012}
Ng K. K.~Y.,  Wong K. W.~K.,  Broadhurst T.,   Li T. G.~F.,  2018, \mn@doi
  [Phys. Rev. D] {10.1103/PhysRevD.97.023012}, 97, 023012

\bibitem[\protect\citeauthoryear{Nightingale, Dye  \& Massey}{Nightingale
  et~al.}{2018}]{Nightingale:2017cdh}
Nightingale J.,  Dye S.,   Massey R.,  2018, \mn@doi [Mon. Not. Roy. Astron.
  Soc.] {10.1093/mnras/sty1264}, 478, 4738

\bibitem[\protect\citeauthoryear{{Oguri}}{{Oguri}}{2018}]{2018MNRAS.480.3842O}
{Oguri} M.,  2018, \mn@doi [\mnras] {10.1093/mnras/sty2145}, \href
  {https://ui.adsabs.harvard.edu/abs/2018MNRAS.480.3842O} {480, 3842}

\bibitem[\protect\citeauthoryear{Oguri}{Oguri}{2019}]{Oguri:2019fix}
Oguri M.,  2019, \mn@doi [Rept. Prog. Phys.] {10.1088/1361-6633/ab4fc5}, 82,
  126901

\bibitem[\protect\citeauthoryear{Pagano, Hannuksela  \& Li}{Pagano
  et~al.}{2020}]{Pagano:2020rwj}
Pagano G.,  Hannuksela O.~A.,   Li T.~G.,  2020, arXiv e-prints, \href
  {https://ui.adsabs.harvard.edu/abs/2020arXiv200612879P} {p. arXiv:2006.12879}

\bibitem[\protect\citeauthoryear{Pang, Calderón~Bustillo, Wang  \& Li}{Pang
  et~al.}{2018}]{Pang:2018hjb}
Pang P. T.~H.,  Calderón~Bustillo J.,  Wang Y.,   Li T. G.~F.,  2018, \mn@doi
  [Phys. Rev.] {10.1103/PhysRevD.98.024019}, D98, 024019

\bibitem[\protect\citeauthoryear{{Pang}, {Hannuksela}, {Dietrich}, {Pagano}  \&
  {Harry}}{{Pang} et~al.}{2020}]{Pang:2020qow}
{Pang} P. T.~H.,  {Hannuksela} O.~A.,  {Dietrich} T.,  {Pagano} G.,   {Harry}
  I.~W.,  2020, \mn@doi [\mnras] {10.1093/mnras/staa1430}, \href
  {https://ui.adsabs.harvard.edu/abs/2020MNRAS.495.3740P} {495, 3740}

\bibitem[\protect\citeauthoryear{{Refsdal}}{{Refsdal}}{1964}]{Refsdal1964}
{Refsdal} S.,  1964, \mn@doi [\mnras] {10.1093/mnras/128.4.307}, \href
  {https://ui.adsabs.harvard.edu/abs/1964MNRAS.128..307R} {128, 307}

\bibitem[\protect\citeauthoryear{Robertson, Smith, Massey, Eke, Jauzac,
  Bianconi  \& Ryczanowski}{Robertson et~al.}{2020}]{Robertson:2020mfh}
Robertson A.,  Smith G.~P.,  Massey R.,  Eke V.,  Jauzac M.,  Bianconi M.,
  Ryczanowski D.,  2020, \mn@doi [Monthly Notices of the Royal Astronomical
  Society] {10.1093/mnras/staa1429}, 495, 3727

\bibitem[\protect\citeauthoryear{Ryczanowski, Smith, Bianconi, Massey,
  Robertson  \& Jauzac}{Ryczanowski et~al.}{2020}]{Ryczanowski:2020mlt}
Ryczanowski D.,  Smith G.~P.,  Bianconi M.,  Massey R.,  Robertson A.,   Jauzac
  M.,  2020, \mn@doi [\mnras] {10.1093/mnras/staa1274}, \href
  {https://ui.adsabs.harvard.edu/abs/2020MNRAS.495.1666R} {495, 1666}

\bibitem[\protect\citeauthoryear{Schutz}{Schutz}{1986}]{Schutz:1986gp}
Schutz B.~F.,  1986, \mn@doi [Nature] {10.1038/323310a0}, 323, 310

\bibitem[\protect\citeauthoryear{Sereno, Jetzer, Sesana  \& Volonteri}{Sereno
  et~al.}{2011}]{Sereno:2011ty}
Sereno M.,  Jetzer P.,  Sesana A.,   Volonteri M.,  2011, \mn@doi [Mon. Not.
  Roy. Astron. Soc.] {10.1111/j.1365-2966.2011.18895.x}, 415, 2773

\bibitem[\protect\citeauthoryear{Seto}{Seto}{2004}]{Seto:2003iw}
Seto N.,  2004, \mn@doi [Phys. Rev. D] {10.1103/PhysRevD.69.022002}, 69, 022002

\bibitem[\protect\citeauthoryear{Smith et~al.}{Smith
  et~al.}{2017}]{Smith:2018gle}
Smith G.~P.,  et~al., 2017, \mn@doi [IAU Symp.] {10.1017/S1743921318003757},
  338, 98

\bibitem[\protect\citeauthoryear{Smith, Jauzac, Veitch, Farr, Massey  \&
  Richard}{Smith et~al.}{2018}]{Smith:2017mqu}
Smith G.~P.,  Jauzac M.,  Veitch J.,  Farr W.~M.,  Massey R.,   Richard J.,
  2018, \mn@doi [Mon. Not. Roy. Astron. Soc.] {10.1093/mnras/sty031}, 475, 3823

\bibitem[\protect\citeauthoryear{Smith, Robertson, Bianconi  \& Jauzac}{Smith
  et~al.}{2019}]{smith2019discovery}
Smith G.~P.,  Robertson A.,  Bianconi M.,   Jauzac M.,  2019, Discovery of
  Strongly-lensed Gravitational Waves - Implications for the LSST Observing
  Strategy (\mn@eprint {arXiv} {1902.05140})

\bibitem[\protect\citeauthoryear{Soares-Santos et~al.}{Soares-Santos
  et~al.}{2019}]{Soares-Santos:2019irc}
Soares-Santos M.,  et~al., 2019, \mn@doi [Astrophys. J.]
  {10.3847/2041-8213/ab14f1}, 876, L7

\bibitem[\protect\citeauthoryear{Somiya}{Somiya}{2012}]{Somiya:2011np}
Somiya K.,  2012, \mn@doi [Class. Quant. Grav.]
  {10.1088/0264-9381/29/12/124007}, 29, 124007

\bibitem[\protect\citeauthoryear{{Springel} et~al.,}{{Springel}
  et~al.}{2005}]{springel2005}
{Springel} V.,  et~al., 2005, \mn@doi [\nat] {10.1038/nature03597}, \href
  {https://ui.adsabs.harvard.edu/abs/2005Natur.435..629S} {435, 629}

\bibitem[\protect\citeauthoryear{Sun \& Fan}{Sun \& Fan}{2019}]{Sun:2019ztn}
Sun D.,  Fan X.,  2019, arXiv preprint arXiv:1911.08268

\bibitem[\protect\citeauthoryear{Suyama}{Suyama}{2020}]{Suyama:2020lbf}
Suyama T.,  2020, \mn@doi [Astrophys. J.] {10.3847/1538-4357/ab8d3f}, 896, 46

\bibitem[\protect\citeauthoryear{Takahashi}{Takahashi}{2017}]{Takahashi:2016jom}
Takahashi R.,  2017, \mn@doi [Astrophys. J.] {10.3847/1538-4357/835/1/103},
  835, 103

\bibitem[\protect\citeauthoryear{Takahashi \& Nakamura}{Takahashi \&
  Nakamura}{2003}]{Takahashi:2003ix}
Takahashi R.,  Nakamura T.,  2003, \mn@doi [Astrophys. J.] {10.1086/377430},
  595, 1039

\bibitem[\protect\citeauthoryear{{Yu}, {Zhang}  \& {Wang}}{{Yu}
  et~al.}{2020}]{Yu:2020agu}
{Yu} H.,  {Zhang} P.,   {Wang} F.-Y.,  2020, \mn@doi [\mnras]
  {10.1093/mnras/staa1952}, \href
  {https://ui.adsabs.harvard.edu/abs/2020MNRAS.497..204Y} {497, 204}

\makeatother
\end{thebibliography}

%%%%%%%%%%%%%%%%%%%%%%%%%%%%%%%%%%%%%%%%%%%%%%%%%%

% Don't change these lines
\bsp	% typesetting comment
\label{lastpage}
\end{document}